\documentstyle[aps,amssymb]{revtex}

\newcommand{\beq}{\begin{equation}}
\newcommand{\eeq}{\end{equation}}
\newcommand{\beqar}{\begin{eqnarray}}
\newcommand{\eeqar}{\end{eqnarray}}

\begin{document}
\twocolumn[\hsize\textwidth\columnwidth\hsize\csname@twocolumnfalse\endcsname
\title{Happier World with More Information}
\author{Yi-Cheng Zhang}
\address{Institut de Physique Th\'eorique,
Universit\'e de Fribourg P\'erolles, Fribourg CH-1700, 
Switzerland}
\date{\today}
\maketitle

\begin{abstract}
We explore the consequences of incomplete information in the stable marriage problem. When information capacity of the participants is increased, more favorable games are created and the quality of the matches are also better. The simple model serves as a metaphor for a world with ever more computing power that leads to more positive inter-personal interactions and of higher quality.
 \end{abstract}

\vskip 1cm
\twocolumn]

\section{Ever More Computing Power}

Computing power has been increasing at a spectacular rate in the last two decades. According to the popular 
Moore's law, it doubles every 18 months. And this trend will not end in a foreseeable future.
At the turn of century, pundits predict future exploitations of this monstrous amount of computing power: long term weather forecast, DNA decoding, and other grandiose projects. In this essay I would emphasize another broad area where this huge computing power can be used for the benefits of the human society. Ever more mutually beneficial inter-personal relations, be they economic or social, are created and improved by increased information. Before I advocate the broader thesis I want to examine in some detail a metaphor model, {\it the Stable Marriage Problem} (SMP).

\section{Stable Marriage Problem Metaphor} 

Gale and Shapley introduced this model as an optimization problem more than three decades ago[1]. Game theorists and economists are familar with it and recently physicists have also shown considerable interest. I refer readers to the previous work[2] for the complete description of the problem as well as more recent applications. Let us consider the simplest version. A world consists of $N$ men and $N$ women. Each has a wish list ranking the individuals of the opposite sex and all the lists are randomly chosen. For example on the wish list of man 1, Claudia is the top choice and Monica the second and so on till the last woman; man 2 has completely independent taste, and his list is a random reshuffling of the $N$ women. But women have also their wish lists. The problem states that $N$ marriages should be established by all the members of this world. Achieving one's top choice is hard because of competition from rivals and your dream partner may not have reciprocal feelings. Recently it has been shown [2] there are about only $N$ possible stable matches, out of $N!$ possible combinations. 

\section{Limited Information Capacity}

The model, like many other examples in economic theory, implicitly assumes that information is infinitely available. To establish a wish list of
$N$ possible partners is no easy matter. A tremendous amount of information is needed to study them
all and to process all the information to establish one's list. In this essay I want to make a crucial
modification, that each member of this world has only limited information processing power, hence
he/she cannot establish a complete wish list. I reexamine the stable marriage problem allowing only limited information capacity to each member. Then I shall let this limit gradually 
increase to see what benefit, if any, increased information can bring to the members of this toy world.

Let us first recall the Gale-Shapley (GS) algorithm. Men propose and women judge. In order for men to
propose, they have to know exactly whom they want in all circumstances: e.g. if man 1 is rejected
the 99th time, he knows whom he should propose to on his 100th try. Women in the GS scenario, on the other hand only decide, upon evaluating proposals. Each woman receives only a finite number of proposals, and this number is on average just one. Since women are {\it passive} players their wish list is superfluous. Such waste is now disallowed if information processing is costly. The same 
GS algorithm is re-interpreted so that women don't have their wish list, being deprived of information processing power. To be politically correct, one notes that the roles of men and women can be inverted. Women nevertheless process the necessary judgemental power when more than one candidates wooing, she can choose among them, given that each time this number is rarely much larger than one.

\section{Mean Field Theory}

Let us next assume that both men and women have only limited information processing power. The complete wish list requires examining $N$ potential partners. Let us assume men can only 
check out a subset of $M$ women and women $K$ men, with the constraint $N\geq M\geq K$.
Each person's subset is chosen randomly from the total population of the opposite sex. In reality no one
knows all other people. We still consider the complete wish list for each, which only God knows and it allows evaluating how well they fare with limited information. We further assume that both men and women
are active players: they each time make one proposal and evaluate possible candidates as well.
Gale-Shapley is an extreme case where women do not have their lists. Another extreme case is when both men and women have complete information, $N=M=K$. For this latter case let us estimate their compromise point $X$, measured as the distance from the top choice. Since both men and women are active players, $X$ must be equal for everybody on average. A man sends out, one each time, a total of 
$X$ applications, before finding a spouse. Each of these $X$ applications has the  probability of
$X/N$ of being accepted, since women have the same pretention as men. Thus the total probability
of acceptance for a person $X^2/N$ must be unity to be matched. This leads to the result $X=\sqrt N$, which is qualitatively correct, as compared to the exact results using more elaborate methods[2]. The above approach is sometimes called `mean field approximation' in physics. In the current case this approximation turns out to be quite accurate, and lends easy generalizations when limited information is considered.

Now a man has only a partial wish list ($M<N$). Without loss of generality we consider these
$M$ women uniformly distributed on his would-be complete list of $N$. This supposes that his list
is sparse, with the average interval of $N/M$. Suppose in a world with 1000 women, but his subset contains only 10. We consider on average that his 10 female acquintances punctuated on his virtual list of 1000. If he is rejected by his first choice, he would propose to his second choice, oblivious of about 100 women better than his second choice on his partial list. Denote by $X$ the 
distance that a man has to go down before he finds his spouse and $Y$ likewise for a woman.
The game starts by each man and woman proposing to someone and evaluating possible proposals received at each time step. Since a woman's list is sparser than that of a man, she goes down her list faster. By the time
all men and women find a suitable spouse they have gone down distances $X$ and $Y$, respectively.
A man sent out a total of $X M/N$ proposals and a woman $Y K/N$, on average. Each proposal from a man has a chance of $Y/N$ being accepted. Therefore $XYM/N^2$ is the probability that a man's proposal being accepted. Likewise, marriages due to a woman's initiatives have the probability
of $XYK/N^2$. The total marriages thus formed can be divided into three
categories: 1) men initiatives only; 2) women initiatives only; 3) mutual initiatives. The first 
is realized when a man proposed and a woman accepted but he was not on her list. The second
is with men and women inverted. The third is due to mutual initiatives when both a man and a woman happen to have each other on their list. The total
probability must be normalized to unity. The normalization relation turns out to be: $XYM/N^2+XYM(1-K/N)/N^2=1$.
The first term is the sum of probabilities of type 1 and type 3 marriages; the second term is
the probability of marriages of type 2 in which a woman proposed to a man but he did not have her on his list, in other words exclusively due to a woman's initiatives. On the other hand, since information is costly men and women go down their lists one step each, we obtain another relation: $XM=YK$. Two relations and as many variables lead us to the solution: $X=N\sqrt M/\sqrt{(M+K-MK/N)K}$ and $Y=N\sqrt K/\sqrt{(M+K-MK/N)M}$. 

\section{More Information Improves the World}
The first conclusion is that $X<Y$, i.e. more information gives an edge to men over women since 
$M>K$. Whoever has more information enjoys better benefit. Now abolish the constraint $M>K$, women are allowed equal or more information than men. We want to know with increased information on the women's side {\it unilaterally}, how the benefits ($X,Y$) change. 

\begin{figure}[h]
  \centering
  \vspace*{70mm}
  \includegraphics{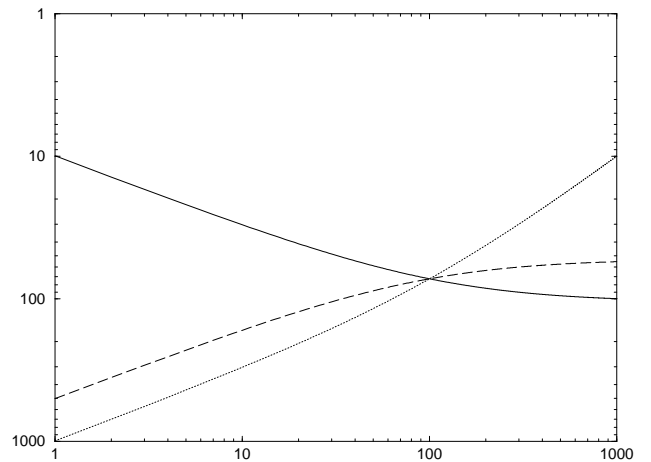}
  \caption{Increasing K while holding M fixed. The vertical
       axis is the partner's ranking. 
       The upper curve (always on the left side) is man's partner's
       ranking, the lowest 
       curve is that of woman, steadly increasing with K. The middle
       curve is the sum 
       of the two rankings, a monotone increasing function of K. }
  \label{fig:1}
\end{figure}

On Fig.1 we plot $X,Y$ against $K$, with $N=1000$ and $M=100$ held constant. We see that when $K<M$, women fare worse than men. As $K$
reaches $M$, the benefits equalize $X=Y$. And as women surpass men in information capacity,
their benefit increases consistently. During this process, men's benefit deterioriates, even though their information capacity does not. However, closer scrutiny reveals that the women's gain is larger than the men's loss. It can be easily checked that the first derivatives of $S=X+Y$ respect to $K$ and $M$ are both negative-definite. This implies that the model is not a zero-sum game.
Increased information, even unilaterally, is beneficial to the society's total.
We may also examine the information asymmetry between men and women. Let the total information capacity be fixed $M+K=const$. We may increase $K$ and at the same time reduce $M$. Of course the benefit to women increases and that to men decreases. What is more interesting is to consider
the total sum, in economics language the total size of the economic pie. In our case the total
size is the sum of $N-X$ of pie-slice for men and $N-Y$ for women.

\begin{figure}[t]
  \centering
  \vspace*{70mm}
  \includegraphics{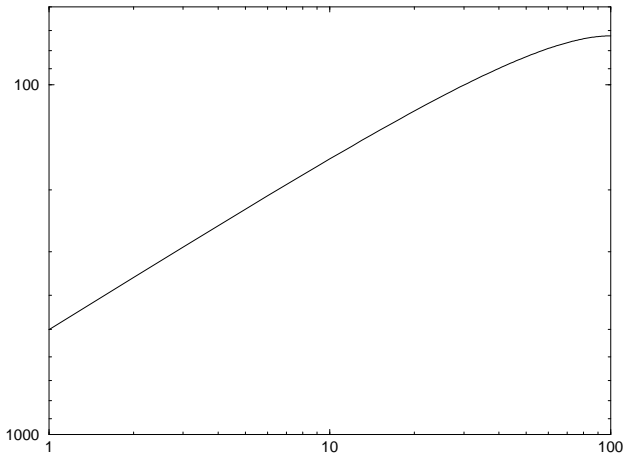}
  \caption{The sum of rankings of man and woman. Total M+K fixed but
  changing K and M. The  
       social optimum is attained with K and M equal.}
  \label{fig:2}
\end{figure}

In Fig.2 we see that the pie's
total size is largest when at the symmetrical point $M=K$. What we can conclude is that the information capacity can bring a considerable edge to a relation which is in principle beneficial to both sides. This benefit is not divided equally. If the pie is divided in the middle, the pie's size is actually the largest. This and other non-zero-sum games are discussed in the chapter {\it Magic Pie}, in
a forthcoming book[3]. We see that unilateral information increase is the least effective way to increase the society's total benefit.
Ideal is the special case that both men and women at equality ($M=K$): we have $X=Y=N/\sqrt{2M-M^2/N}$.

We may ask if there is unequal information capacities between two groups initially and capacities increase
simultaneously on both sides. Would the two groups' relative benefits stay the same? This is a relevant question since if there is some initial inequality, it is more natural that information capacity would rain down on them indiscriminately. We can answer this question using our above model. Benefits to men is $N-X$ and that of women $N-Y$. Now assume the information capacities increase by the same factor: in the places of $M$ and $K$ we have $\alpha M$ and $\alpha K$, where $\alpha$ is a constant larger than unity. The relative benefits can be defined
as the ratio, $R=(N-Y)/(N-X)$. Assuming $K<M$ initially, it can be verified that $R$ increases monotonically as $\alpha$. Thus we can conclude that within our model the weaker side's
relative benefits increase with unifromly increasing information capacities. This is important since the uniform increase is probably the most realistic way. It's comforting to know that the weaker side improves faster than the stronger side.

\section{Risk of Being Left Alone}

Next, consider that our men and women are not so easily pleased. There is a threshold below which they would prefer to stay alone rather than marry. Denote by $\eta$ ($0\leq \eta\leq 1$) the threshold value. $0$ means accept-all and $1$ reject-all. For example, a woman with $\eta=0.9$
would find only 10 men {\it acceptable} out of a random pool of 100 men, on average. It is quite realistic that a person find only a
small fraction from a random pool of the opposite sex acceptable. If all those in this small subset do not reciprocate, then she or he would stay alone.

With limited information capacity men and  women may first encounter the barrier of $\eta$, before going down distances of $X$ and $Y$. Let us use re-scaled parameters $x=X/N$ and $y=Y/N$.
The bottom lines for men and women are $x_{cut}={\rm min}(x,1-\eta)$ and $y_{cut}={\rm min}(y,1-\eta)$, respectively. Therefore if somebody hits her/his bottomline before the required distance, she or 
he risks remaining a bachelor. The number of bachelors in such a world is $B=N\max(0,(x_{cut}+\eta-1)/x_{cut})$ which is a decreasing function with increasing $M$. 
For simplicity we consider symmetric case that the information capacity is the same for both men and women $M=K$. If all the members have too high expectations of their potential mates, some
would remain single even with complete information $M=N$. In modern societies, anecdotal
evidence suggests that people's expectations are slowly rising, concomittantly with increased 
information capacity. It remains to be seen if information increase can outrun the ever more demanding wish lists. 

Consider the total benefit that increased information brings to such a society. The benefit for a person is defined as a linear function (avoiding te need of utility function) of the distance that her/his spouse measured from the
top choice. In other words, the shorter the distance the happier he/she is. Denote the function
by $g(x)=(1-\eta-x)/(1-\eta)$. This linear (in $x$) function is calibrated to the highest value
unity when the top choice ($x=0$) is taken and zero at just the threshold ($x=1-\eta$) and negative
below the threshold. The total benefit to the society is then the integral
$ G(M)=N\int_0^{x_{cut}}g(x) $
for the symmetric case.

\begin{figure}[t]
  \centering
  \vspace*{70mm}
  \includegraphics{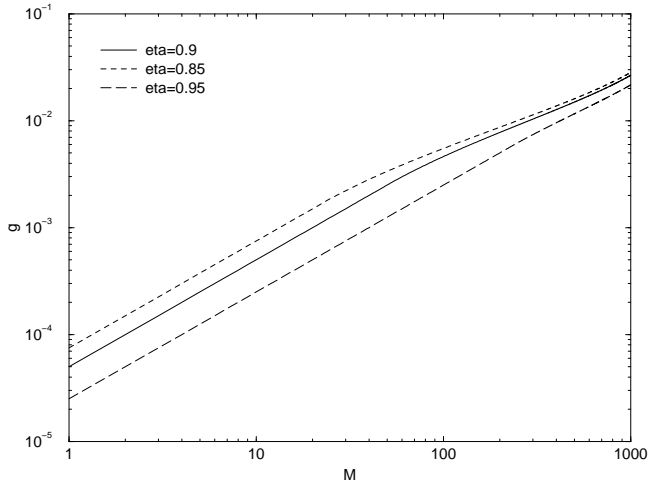}
  \caption{Global gain with increasing information capacity M for various threshold values.}
  \label{fig:3}
\end{figure}

 We see that $G(M)$ shown in Fig.3 is a monotonically increasing function, which is actually two smooth functions pasted together at $M^*$, where all the bachelors just barely find a suitable spouse. When everyone is married, the benefit continues to increase since on the aggregate level the distances from the top choices decrease as more suitable couples are formed.

This is not strictly Pareto-efficient since somebody may fare worse in the information rich region
(large $M$). Somebody may have a spouse in an information poor region (small $M$) and increased information makes his/her mate run away with somebody else, and he/she has to settle with another
mate of lower ranking. Before he/she was enjoying the benefit of `ignorance' of the spouse. More information enabled spouse now knows better her/his own available options. However, on the aggregate, the total benefit to the society increases steadily. Thus a minority of members actually suffer whereas a majority benefit from enhanced information capacity, even when the increase is uniform for everyone. The fate of a particular person's welfare can be studied in details and will be discussed elsewhere.

The above discussion may have direct applications in modeling literally real marriage and dating. Powerful computers may allow more people access their potential mates in
more and  more efficient ways. One should interpret the information capacity broadly: computers
also indirectly power tele-communications, wireless; airlines ferrying people to meeting places
need also computers; peers of interest groups can have wider reaches with the internet etc. 

\section{Matching Consumers and Products and Beyond}

However my aim here is to show that the implications go far beyond marriage, in many other
relationships in the human society. The marriage model serves but as a metaphor for broader applications. The most relevant relationship is commercial. 
Consider the relationship between firms and consumers. It is generally assumed that a fair transaction 
between a firm with a product and a consumer needing that product is beneficial to both parties.
In economics text books it is asserted that the benefit to both sides are {\it equal} and 
the so-called economic pie is the largest. But modern products and services are complex and ever more so in the future.
For the compatible consumers and firms to find each other and it takes time and effort. The simple law of supply and demand based on price alone is not adequate any more. We never know all the 
potentially useful products and services available in the world. Likewise, a firm can never guess right everything consumers want. An essential part of a commercial transaction lies in searching for each other. 

Borrowing from the above marriage metaphor we see that the economic pie is the largest when information capacity is equal on both sides. However, it is not hard to see that in most
commercial transactions there is the perennial information asymmetry between the consumers
 and firms. This doesn't necessarily imply that consumers are stupid or information challenged. This has more to do with the fact that for most products/services, consumers care less than firms
producing them. What is a marginal business to a consumer is generally a core business to
the firm specialized in producing it. With this asymmetry, the economic pie is not largest, but could be made larger with the advent of the internet and huge computing power. More on this (in)efficiency and how much information asymmetry can be restored are dealt with in a forthcoming book[3].

Commercial transactions differ from marriage in some important aspects: a consumer needs multiple products/services simultaneously. In marriage language this would correspond to polygamy and promiscuity.
While marriage is done once and for all, a commercial relationship is repeated and constant search. A restaurant you visited last year needs to be checked again for its current status. 

Human relations are not limited to marriage and commercial transactions. In a broad sense each individual relies more and more on others, especially her/his peers. In the internet age, peers
of a certain activity located oceans away can find each other. Human creativity can be boosted by 
knowing potential demands. Economic institutions and knowledge institutions become more and  more elaborate, permitting previously unthinkable cooperations to be realized, to the total benefit of the human society.

To have an idea what sort of human relations (noncommercial) can be matched, think of a plane-load passengers filled in a narrow space on a trans-atlantic flight. Currently seating is random in the sense hardly any consideration is given as who your sitting neighbors are. Suppose in the near future there will be such a service: an airline can arrange sitting matches according to some sort of affinity of interests, just like in marriage. In a recently work (PRL) Maslov and I have formulated a generalized information theory giving a framework as how the taste
space can be coded and suitable matches predicted, based on only a small fraction of observed data. Differing from marriage is here that sitting matches are only for a short time span, and new matches have to be arranged over and over again. 

Why passengers would want to be matched?
Let's say a random pair of passengers assigned to a narrow space is slightly detrimental to both.
If one does not agree with this assertion, it suffices to observe in a public park there are
some empty and occupied benches. A newcomer would more likely to choose an empty bench than to sit next to a stranger. However, in the real world there are potential positive pairs but they are rare. That's why before we have any knowledge about the next person our attitude is to assume that the average value of the matching between ourselves and that stranger is somewhat negative. For the sake of arguments let's assume that among the pairwise links of the 6 billion people $80\%$ of such potential matches are negative, $20\%$ positive. Among the $20\%$ positive matches $10\%$ are very 
stimulating, $5\%$ of the $10\%$ exciting, $2\%$ of the $5\%$ fantastic, and the top $1\%$ has life-changing 
potential (e.g. a marriage, new job, life-long friendship). In short positive matches are
rare so lengthy computation is needed to dig them out. Why an airline is interested in providing
such a service? In any case passengers are matched by random chance, it's sensible to provide something with the potential of better customer satisfaction. Even though such matches are not commercial,
they certainly can help an airline's bottomline indirectly.

\section{Self-Searching or Matchmaker?}

Next we ask the question how do people find each other. There are two basic modes: 1) everybody searches for himself; 2) searches are done through a matchmaker. Fig.4
below gives a caricature of the idea. In the self-searching mode each enquires about all others, so there are about $N^2$ enquiries. 

In other words the society spends $N^2$ units of effort. Matchmaker-mediated searches are much more economical, necessiting only $N$ units of effort. Since information is always
finite, as we shall see that our problems of matching the world will be more and more complex.
Beside the obvious economy of the matchmaker-mode, there are some general pros and cons of the
each alternative. 

\begin{figure}[t]
  \centering
  \vspace*{60mm}
  \includegraphics{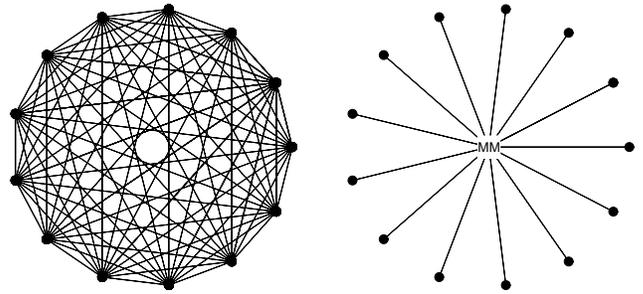}
  \caption{Self-searching mode versus matchmaker-searching mode, the latter has obvious economy 
       of effort.}
  \label{fig:4}
\end{figure}

Besides economy on effort, a matchmaker would try to make the largest social pie as her goal,
i.e. the sum of the benefits of all the members, minus charges. As we learn from the SMP that in the self-search mode everybody strives to improve her own happiness as much as she can, without
any consideration of others. Therefore the stability is important in determining the final matching. In the illustration Fig.5 below, a miniature model of two men and women located in a one-dimensional space.

\begin{figure}[t]
  \centering
  \vspace*{64mm}
  \includegraphics{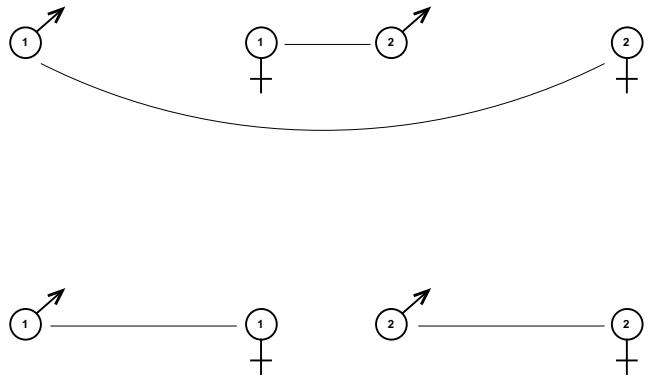}
  \caption{In the upper panel when man 2 and woman 1 happily married
  and man 1 and woman 2 very unhappy; 
       if man 2 and woman 1 were considerate by agreeing 
to `sacrify' a little,
the total social
benefit is greatly enhanced (sum of distances is smaller).
}
  \label{fig:5}
\end{figure}

The attractiveness is defined as the separation distance between a man and a woman. The closer a man to a woman, more attractive they are to each other.
If we let them do self-searches, man 2 and woman 1 would like to be matched, and man 1 and woman 2 have no alternative but to accept the bad match. The total social costs equal the sum
of the two lengths, which is high in this case. On the other hand, if man 2 and/or woman 1 were slightly considerate they would realize that by sacrificing a little taking their second best choice,
the social welfare would be greatly improved, as shown in the second panel. However, they would 
not care their fellow citizens's fate without a propre incentive and institution. The
matchmaker has the social welfare on her mind. Not because of magnanimity, but her profits derive ultimately from the membership fees, which are in proportion of the total social pie. This problem can be handled by a game theory model similar to that of John Maynard-Smith, where he typically treats two species of selfish players. Achieving
generalization to games of {\it menage a trois} is straightforward but lengthy, will be presented
elsewhere. It can be shown that two groups with partially opposing interests (like our men and women in the SMP model), plus the selfish matchmaker, can achieve superior social pie and yet stable since each strives to do well for herself/himself, while inadvently doing good for the society.

Let's go back to our SMP model. In a world of 10000 men and 10000 women with only partial information. The ideal limit
with infinite information can be solved [2], both for self-searching and matchmaker-searching modes.
In the self-searching mode, a person can expect to find a mate ranked at 100, out of her full wish list of 10000.  In the matchmaker-searching mode, a person
can expect to find a mate ranked 81 on her full wish list, a significant improvement taken into account that everybody in the population gets that much on average. The improvement in rank by 19 is the matchmaker's edge. In principle it's more advantageous for the men and women to join the matchmaker's service. When information capability is limited, the matchmaker's edge is larger, as shown in Fig.6. 

\begin{figure}[t]
  \centering
  \vspace*{65mm}
  \includegraphics{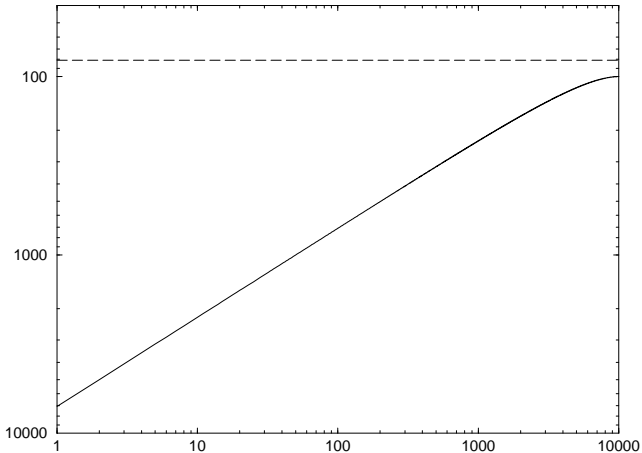}
  \caption{Matchmaker has the possibility arranging matches in ground
  state, which is the upper curve. 
       Self-searching mode, with partial information fares much less
  well, including the special case 
       of complete information.}
  \label{fig:6}
\end{figure}

If the total information capability of the society is capped by a fixed amount, it's more effective to let the matchmaker use it. In the real world it's likely
that both modes coexist, at least for a long time before the superior matchmaker institutions
become dominant.
We see that in general there can be conflicts between the solution by a matchmaker and that by individuals themselves. This is a fundamental conflict since in a viable society one should not 
count on people's selfishless deeds to sustain it. Our matchmaker acts as if she were the proverbial benevolent social planner, exists only in pure theory. 

However, it is not that outlandish
to suggest that such social-optimal solution can indeed exist, even each individual is selfish.
We mentioned above that if the frequency of the matches is high (marriages are least frequent examples, supposedly to happen only once in a lifetime), a selfish individual may forego some of
his immediate gains, in hoping to be compensated later. His reasoning is quite rational: since in frequent matches he can end up on the sucker's side as often as on the gainer's side. If such an institution (matchmaker in our example) exists keeping record of good deeds and being able to
easily enforce the fair-play rule, then it's to his own advantage to respect the matchmaker's ruling, even occasionally he's summoned to sacrify. The overall benefits from such an institution is positive since we have nonzero sum games here. Then this begs the question how often is often enough? 

To get a clue we have to face the reality that an individual's immediate
gains are always more important than future hypothetical debt-repaying, even with the knowledge
of effective enforcing is in place. We can model good deeds by proposing a memory effect: a credit that is done now decays with time. The decay can be a power law, or an exponential one,
$\exp (-t/t_0)$. $t_0$ is the time scale that beyond which the memory about a good deed fades and is no longer counted. If somebody offered you a cup of coffee 10 years ago you may not keep that in your gratitude-to-return list. Shorter is this scale, easier for the matchmaker to enforce the fair game. In the information age possible matches will be ubiquitous and most of them have 
a high frequency. We should expect that many future matchmakers can in fact act as a benevolent planner. There will be a {\it phase transition}: when the frequency surpasses a given threshold, suddenly the matchmaking mode becomes viable. The society may not respond right away,
but the superiority of the matchmaker mode is hard to resist, just like a super-cooled liquid below the freezing point. 

As an example of higher frequency matchmaking, let us consider the much hyped third generation wireless communication. The industry is making a trillion dollar bet
on the always-on wireless communication mode. From a PC to a always-on wireless device, the frenquecy of being connected is much higher. The industry actually has not figured out what to do with such a high communication frequency (not to be confused with the technical synonym). Proposals are abundant, people might want to check frequently their stock trading, or to keep track their online casino bets, or to digging for more music feeds, etc. However, we cannot expect most people to become day traders, to be gambling addicts. The wireless future rests on 
the exploitation of mediating inter-personal relations, with whatever activities a person does
offline already. This will be appropriately dealt with in the afore-mentioned book.

\section {Imperfect Information} 
In a previous section we have introduced the concept of limited information, that a player can only access a subset of all the available alternatives. We need now to address another source of limited information capabilities, i.e. the information obtained in reality is imperfect. When an agent obtains information about a subject, his knowledge is only an approximation of the truth.
Consider a consumer searching for products, she may have the capabilities checking on a large number of potentially useful products, but her results can only be a distorted picture about the
real nature. The perfect truth is unknowable. Though better efforts can lead to better precision in general. We can model imperfect information through a simple example. Let there be $N$ (even) $M$-dimensional binary vectors $V_i, i=1,...,N$. Our task is a standard matching problem where the vectors have to be pairwisely matched, and the sum of the all the pairwise overlaps has to be minimized. However, now we have to face a new complication: we cannot determine the exact nature of each bit. If a bit is $0$, we have the possibility of detecting it both ways: $0,1$.
The truth occurs with probability $p$ and the error with $1-p$. First consider the case $2^M>N$, i.e. rarely two vectors would be identical in a given sample.  Even with perfect information,
complete overlap is impossible to achieve, for the all the vectors, the best match one can obtain is $M'\approx M/2+\sqrt{M/2 \log N}<M$. Where $M'$ denotes the average best partial overlap. Now consider imperfect information, and for simplicity we assume $2^M<N$, so in principle with perfect information ideal overlaps are obtainable $M'=M$. Since now the precision
is limited, the best one can hope is the overlap $M'=Mp$. 

One important conclusion emerges that the quality of the matches grows with the number of the potential pool, but this growth stops at an intrinsic size $N_{\rm max}$. The total system of 
$N$ players (here represented by $N$ vectors) appears to be divided into subsystems of size $N_{\rm max}$, which turns out to be proportional to $\sqrt M \exp {(2p-1)^2 /M}$. As the precision
(i.e. $p$) increases, we see that $N_{\rm max}$ increases exponentially with $p$. What this analysis means is the following: when the information is very imprecise (small $p$), the total system breaks into smaller groups, since intrinsically there is no need to search better matching partners in a larger pool. As precision improves, larger pool brings better matching quality. This is analogous to the correlation length in physics, the precision $p$ plays the similar role of inverse of temperature. At high temperatures (small $p$), a physical system loses coherence over distances larger than the correlation length. As temperatures become lower,
the coherence region grows and even become divergent. In real life matching example can observed
almost everywhere. In a primitive society where information about anything is very imprecise, therefore the economic activities are confined to limited and isolated groups, with no added advantage of cross-group contacts. As information becomes abundant and more precise, the connected region becomes also larger, eventually leads to a global economy.

\section{Law of indeterminacy}
Lest I give the wrong impression the searching and matching problem in economic and social relation is reduced to an optimization problem, as hitherto is the founding paradigm in neo-classical economics. Human beings are different from inanimate objects in that, under searching and matching action, they react. Depends how effective a searching is done an individual's quality (what she can offer) as well as the demand (her wants), at the same time evolve as the more powerful and effective searching is being carried out. Economic literature treats these variables as if as given, immutable objects. If information is very imprecise and primitive, an individual would not be stimulated enough to bring out of her latent qualities, since she is aware of that it wouldn't be appreciated enough anyway. Most these talents need time and effort to reach their prime, even in best circumstances. Likewise, an individual would not wish too much in a information-challenged society, being forced to be realistic. And tastes, even the most frivolous, need time to develop as well. Send a Santa Claus to the famine-stricken sub-sahara to offer people whatever they {\it want}, few would clamour for a {\it Ferrari}, as a Wall Street trader would.

It's not hard to model the above idea modifying the matching problem. In the standard matching problem described in the last section, we have a Hamiltonian $H$ which is the sum of all the pairwise overlaps of among the $N$ vectors. The Boltzmann weight is proportional to $\exp-H/T$, where $T$ is the temperature. Smaller is the temperature, better matches would be found since the rare combinations now receive prominent weights. When $T$ approaches to zero, the standard
matching problem becomes an optimization problem, and in the past years have been extensively
studied by Parisi and his co-workers. In our problem of economic and social matching, there is 
a natural interpretation of the temperature: more powerful searches correspond to lower temperatures, and in fact we may postulate a relation between the society's "temperature" and the current computing power. So $T$ can be thought as a function of the aggregate computing power $C$: $T=T(C)$. It is conceivable some simple power laws $T=1/C^\alpha$ is operative, as some preliminary analyses show.

The key consideration not to fall into the optimization trap is to postulate that human beings
react to the ever more powerful searching and matching. The model is quite simple, the complexity of an individual is specified by the dimensionality of the vectors, $M$. We need to relax the usual constraint that $M$ being a constant. Under the reduced temperature (more computing power), $M$ is likely to increase. We may further suppose that $M$ is a certain function of $T$: $M=M(T)$, the smaller is $T$, the larger $M$. We shall see that under the current "global cooling" tendency, at least as far as computing power evolution is concerned, our optimization tasks get more and more difficult. It's like chasing a running objective and there is no fixed target to aim, as prescribed in neoclassical economics. Details about the model and broader discussion will be presented elsewhere. The most interesting is to note from such a runaway optimization problem is that whether the ever more powerful computing power is
able to catch up the ever more complex problem. This depends the choice of $M(T)$, which ultimately depends on real experimental findings. The most likely scenario would be that the problem remains just marginally intractable. Nature reserves enough wisdom to tease the technically confident people in the pursuit of `optimum'. So much is the essence of the law of indeterminacy.

\section{Pie-Slicing, Matchmaker's Dilemma}

It is instructive to consider a toy model of matchmaking, to illustrate how a selfish matchmaker
can do well while doing (partial) good for the society. Consider a club of many members to be matched and a matchmaker-entrepreneur. The total membership $M=M(b)$ depends on the benefit per member $b$, a supposed given function. Denote $a=a(V)$ the attractiveness to
a single member, $V$ is the total investment that the matchmaker plows back to make the whole
game attractive. The individual benefit $b=(1-\gamma)a$, i.e. the attractiveness received minus the membership fee $f=\gamma a$ ($0\leq\gamma<1$). The total membership fees collected by the matchmaker is $\gamma M a$, a $g$ fraction is kept for profit $G=g\gamma Ma$, the fraction $1-g$
is plowed back as investment $V=(1-g)\gamma Ma$. This is a self-sustained, static model and
can be solved readily. The solution can be expressed as functions of $\gamma$ and $g$. The total membership $M=M(\gamma, g)$ and attractiveness $a=a(\gamma,g)$, the total social pie $Z=Z(g,\gamma)=Ma$, before profits and investment; the total membership benefits $B=(1-\gamma)Z$;
and the profit for the matchmaker $G=\gamma g Z$. If the matchmaker has her way, she would choose
$\gamma$ and $g$ in such a way to maximize her profit $G$; if the members have their way, they
want to choose $\gamma$ and $g$ to maximize their benefit $B$; finally in a utopia a benevolent planner would like to see the largest possible social pie, he would want to see the total gross pie being largest. Let us limit ourselves to a concrete example: $a(V)=(V/V_0)^{1/3}$ and $M(b)=M_0 b^{1/2}$. The utopian largest pie
$Z_{\rm max}=1/4 {M_0}^2/V_0$ can be achieved with $\gamma=1/2$, $g=0$. The membership benefits
maximizing solution gives the gross social pie $Z_{\rm M}=8/9 Z_{\rm max}$ with $\gamma=1/3, g=0$. The matchmaker's maximal profit solution with $\gamma=2/3, g=1/2$ makes the total social pie considerably smaller $Z_g=4/9 Z_{\rm max}$.

\begin{figure}[t]
  \centering
  \vspace*{45mm}
  \includegraphics{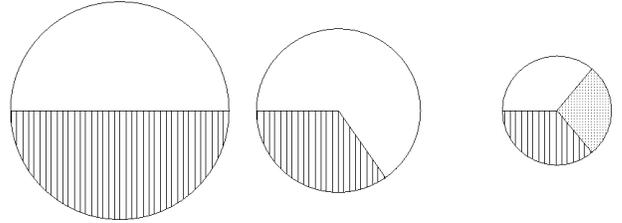}
  \caption{First panel shows the utopian social planner can mediate
  the largest pie where investment is half 
       of the pie; the middle panel is what the members wish for, only
  1/3 goes to investment; and selfish 
       matchmaker maximal profit is the realistic outcome, though the
  pie is the smallest, but still better  
       than self-searching.}
  \label{fig:7}
\end{figure}

In Fig.7 the panels show the sizes of the gross social pie, as well as slice-shares. It can be shown that self-searching mode achieves much less a social pie, here the membership fee is replaced by a person's own search effort costs. 

We see that the members and the matchmaker have partial interest overlap, for that they can be said to live in a
symbiosis. They are mutually interested in having a large pie. But this overlap is limited and
while the members want the solution in the second panel, the greedy matchmaker is interested in having the solution in the third panel. Denote $\Sigma=B+G$, start from the second panel and hold the investment fraction $\gamma (1-g)$ fixed and increase $\gamma$ and $g$ to transit to the solution in the third panel. As a matter of fact in this range it is a negative-sum game between the matchmaker and the members since $\Sigma$ decreases monotonically. The matchmaker would be less greedy as soon
as there is competition. Competition among matchmakers is somewhat different from firms selling same goods since information is a very pecular goods. This goes beyond the scope of this
essay. 

\section{Double-String for Taste Space, or Social DNA}

There are many human activities that would benefit from effective matching. However some activities
are easier to predict and mediate than others. For example, if you like neither hot pizza nor hot soup, it'd be a rather good prediction that you don't like all other hot food. We need a systematic way to model people's tastes and their affinity. It's clear that tastes are multi-dimensional, for simplicity we code them by bit-sequences with each entry representing absence or presence of a given attribute. Our aim is to match people with people or to things.
However, a single bit-sequence is not enough for a person, as his wants and his qualities judged
by others are in general not the same. A short woman may prefer a tall man, a blond man with blue eyes may prefer a red haired and green
eyed woman. In the simplest model a person can be characterized by two bit-sequences: one for what
she has to offer $Q=(q_i, i=1,m)$ (Quality), another what she wants $W=(w_i, i=1,m)$ from others. When a man is matched to a woman, the happiness of the man is defined as $H=(1-\epsilon)h(x)+\epsilon f(q_1w_1,q_2w_2,...,q_mw_m)$.
where $x=q_1w_1+...+q_mw_m$ is the overlap, where $Q$ is the quality-vector for the woman and $W$ wants-vector for the man. The happiness for the woman can be likewise defined with appropriate labels. $h(x)$ is a single variable function, whereas $f$
is a random boolean function of m bit-variables. For $\epsilon=0$ there is a lot of coherence among the preferences
in the population. For instance the top two choices of man 1 would end up in another man's list 
close to each other. In other words the entropy is much less than that in the ranking case. For $\epsilon
=1$, on the other hand, everybody's taste is completely idiosyncratic. In the above example, the top two choices would end up in arbitrary places in other men's list. If $\epsilon$ is small,
there is a quite good chance of predicting peers' tastes, without knowing everybody in the population. This is called affinity. On the other hand, for $\epsilon$ large tastes are
very unpredictable and full knowledge of each individual's tastes is needed before a matchmaker can
mediate. In reality we don't know {\it a priori} whether the affairs are easy to predict or not.
To decide, I propose a simple criterion. Take a person's top two choices, find the distance of these two choices in another member's list. Average this distance over all other members and repeat the
same procedure by considering all members' top two choices. Call this parameter $\eta$. It can
be readily shown that $\eta=\epsilon/2$ to the leading order approximation, using the above ansatz. An alternative measure would need to calculate the full entropy.

\section{Outlook}

Let's end this essay by speculating a bit ahead where the current information revolution would lead us. If my view is only partially correct: A small but increasing fraction of future computing power will go to mediating human relations. With the increased information capacity each member of the society is made better aware of her available options. As a result more and more positive-sum games are formed and their aggregate quality is also improved.  The information age can be roughly divided into two stages: 1) in the first stage that is about the last twenty years before the turn of the century, {\it people make computers connected} with the advent of internet etc. This stage is by no means ending as grid-computing is just about to start. 2) In the second stage {\it computers help people connected}, in small measures this has already started.
In fact the ever more powerful computing power pouring into the society matching people together to their mutual benefit. As information searching and computation power is increased the entropy of the world in human resources in further reduced, somehow a compensation for the entropy loss (increase) of the physical world. Indeed, more computing power could make a world smaller and happier.

Acknowledgements: I've benefitted from conversations with A. Capocci, P. Lauretti and S. Maslov.

\end{document}